 \documentclass[12pt]{article}

\newcommand{\ihat}{\hat{\textbf{\i}}}
\newcommand{\jhat}{\hat{\textbf{\j}}}
\newcommand{\khat}{\hat{\mathbf{k}}}

\usepackage{epsfig}

\usepackage{amssymb}

\usepackage[ps2pdf,%
a4paper=true,%
breaklinks=true,%
colorlinks=true,%
pdfauthor={First Author et al.},%
pdftitle={Template for manuscripts in Advances in Space Research}%
]{hyperref}
\usepackage{graphicx}
\usepackage{amsmath}
\usepackage{url}

\title{The flyby anomaly:\\ A case for strong gravitomagnetism ?}

\author{L. Acedo\thanks{The author gratefully acknowledges the JPL's Horizons system team for the web-based software used in the calculation of the ephemeris in this work.}\\
Instituto Universitario de Matem\'atica Multidisciplinar \\
Building 8G, $2^{\mathrm{o}}$ Floor, Camino de Vera,  \\
Universitat Polit$\grave{\mbox{e}}$cnica de Val$\grave{\mbox{e}}$ncia, 46022, Valencia, Spain\\
e-mail: luiacrod@imm.upv.es}

\begin{document}
\maketitle






\begin{abstract}
In the last two decades an anomalous variation in the asymptotic velocity of spacecraft performing a flyby manoeuvre around Earth has been discovered through careful Doppler tracking and orbital analysis. No viable hypothesis for a conventional explanation of this effect has been proposed and its origin remains unexplained. In this paper we discuss a strong transversal component of the gravitomagnetic field as a possible source of the flyby anomaly. We show that the perturbations induced by such a field could fit the anomalies both in sign and order of magnitude. But, although the secular contributions to the Gravity Probe B experimental results and the Lense-Thirring effect in geodynamics satellites can be made null, the detailed orbital evolution is easily in conflict with such an enhanced gravitomagnetic effect.
\end{abstract}


\parindent=0.5 cm

\section{Introduction}
\label{intro}
In the last decades we have entered an era of high-precision astrodynamics in which spacecraft, planets and satellites can be tracked with radar ranging, laser ranging and Doppler techniques \cite{Radar,Laser,Doppler}. These improved methods have
allowed the testing of orbital trajectories in General Relativity to unprecedented accuracy. In particular, the determination of the $\beta$ and $\gamma$ parameters in parametrized post-Newtonian dynamics is now possible to the level of $\sim 10^{-5}$, ruling out alternative models, such as Brans-Dicke theory, or putting stringent constraints upon them \cite{BDPPN}.

Error checking and detailed modelling of spacecraft geometry and thermal properties has also proven useful in the improvement of accurate navigation methods. A finite-element model for the anisotropic thermal radiation from the on-board radioisotope generators has been build for the Pioneer spacecraft. This model has explained away the anomalous recoil acceleration directed
toward the Sun which has been found both for Pioneer 10 and Pioneer 11 spacecraft \cite{PioneerI,PioneerII}.

Another anomalous behaviour of spacecraft was reported in 2008 by Anderson et al \cite{FlybyPRL}. These authors have analyzed the data for several Earth flybys of deep-space missions that took place between December 1990 and September 2005. Flybys are a common manoeuvre in spacecraft missions which allows the spacecraft to gain or lose of heliocentric energy with the purpose of reaching their objective \cite{slingshot}. An analysis of the data for these flybys have shown X-band Doppler residuals that are interpreted in terms of a change
of the hyperbolic excess velocity, $V_\infty$, of a few mm/s. Anderson et al. have proposed the phenomenological formula:
\begin{equation}
\label{flybyform}
\displaystyle\frac{\Delta V_\infty}{V_\infty}= K (\cos \delta_i-\cos \delta_o) \; ,
\end{equation}
where $\delta_i$, $\delta_o$ are the declinations for the incoming and outgoing osculating velocity vectors and $K$ is a constant. The value of $K$ seems to be close to $2 \omega_E R_E /c$, where $\omega_E$ is the angular rotational velocity of the Earth, $R_E$ is the Earth radius and $c$ is the speed of light. Although this formula works reasonably well for the six flybys studied in the paper, the proposal for the relation of $K$ with the Earth's tangential velocity at the Equator is an unjustified and speculative hypothesis, taking into account that the flybys of other planets with different rotational velocities and radii have not been considered.

In the aforementioned paper, Rievers et al. \cite{PioneerI} also analyzed the thermal radiation pressure on a model of the Rosetta spacecraft and found it insufficiently small to account for the anomaly as well as wrong in sign. Other conventional effects that have been considered and estimated are: atmospheric drag, ocean or solid Earth tides, charge and magnetic moment of the spacecraft, Earth albedo, Solar wind and spin-rotation coupling \cite{Lammerzahl}. Iorio has also computed the effect of the perturbation induced by the Earth's gravitomagnetic field as predicted by General Relativity \cite{Ioriogm}. This effect is five orders of magnitude smaller than the observed velocity change.
We will see that the enhanced gravitomagnetic effect, proposed from a phenomenological point of view in this paper, is modulated by a parameter $\beta$ much larger than the prefactor $r_S/r$ (the quotient of the Schwarzschild radius and the distance from the planet's center to the spacecraft) of the standard gravitomagnetism of a rotating planet in General Relativity. Another attempt to explain the anomaly uses the transversal Doppler effect in Special Relativity \cite{Mbelek} but this explanation is also
dismissed because the spacecraft movement is not always transversal to the Earth surface and the anomaly appears also in the ranging data as well as in the Doppler data. It has also been suggested that a second-order error in the integration method could
manifest as the alleged anomaly \cite{Acedo}. 

Moreover, a mismodelling of the coordinate systems may also be involved in the anomaly because the fitting equation of Anderson et al. \cite{FlybyPRL} involves only geometrical parameters \cite{Review2011}. The subtleties of the energy transfer process during planetary flybys have also been analyzed in detail for some real missions in the hope of gaining some insight on the origin of the flyby anomaly but to no avail \cite{Anderson2007}. In the flyby manoeuvres the spacecraft gains or lose energy with respect to the barycenter of the Solar System but the strange fact about the anomaly is the apparent energy change observed in the Earth-centered system. In this paper we will study transient variations of the perturbed velocities and positions for the spacecraft as the consequence of a two-body interaction among the spacecraft and a phenomenological proposal for a gravitomagnetic field of the Earth. So, our results are expected not to be affected by the definition of the coordinate frame.

Some proposals for a non-conventional explanation have also been made: Adler has worked out a model for an halo of dark matter particles bound by the Earth gravitational field. The explanation of the anomaly requires very strong constraints on the cross sections, the energy of the particles and the dominance of inelastic and elastic processes along different trajectories to account for the increase or decrease of the spacecraft velocities \cite{haloDM}. A five dimensional theory of gravity has also been considered by Gerrard and Sumner \cite{fivedim}. Other proposals include radical modifications of well established principles such as the principle of equivalence \cite{inertiamod} or Lorenz invariance \cite{lightspeed}. Busack has also derived a phenomenological formula for a velocity dependent potential with three adjustable parameters which fits the observed discrepancies in the flybys \cite{Busack}. However, these radical proposals are unlikely to receive further support because they are disconnected from the theoretical paradigms or other domains of empirical data.

In this paper we take the data for the flyby anomalies at face value and we consider a scenario in which the anomalies are the
consequence of a modelling force unpredicted by the current theory of gravitation. To select this model we follow an intuition given by the authors of the seminal paper on this topic \cite{FlybyPRL}. In their work, Anderson et al. suggest that the latitude dependence of the anomaly could be related with a frame-dragging effect much larger than the one predicted by standard General Relativity \cite{IorioReview}. It is well-known that linearized General Relativity can be interpreted in terms of a gravitoelectric field (containing Newton's law) and a gravitomagnetic field \cite{Rindler}, which for a rotating planet takes the form:
\begin{equation}
\label{Bgm}
\mathbf{B}=-\displaystyle\frac{1}{5} \displaystyle\frac{r_S}{r} \left(\displaystyle\frac{R}{r}\right)^2 \left[ \boldsymbol\Omega-3\left(\boldsymbol\Omega\cdot\mathbf{\hat{r}}\right)\mathbf{\hat{r}}\right]\; ,
\end{equation}
where $r_S= 2 G M/c^2$ is the Schwarzschild radius of the body, $R$ is the planetary radius and $\boldsymbol\Omega$ is the angular velocity vector. This field imparts an acceleration in a test particle given by a Lorentz force law:
\begin{equation}
\mathbf{a}=-\displaystyle\frac{2}{c} \mathbf{v}\times \mathbf{B} \; ,
\end{equation}
which justifies the denomination of $\mathbf{B}$ as gravitomagnetic field. This effect, jointly with direct detection
of gravitational waves, is one the most feeble and difficult predictions of General Relativity to be measured but, after 
several decades of planning, the gravitomagnetic field of Earth (as well as the larger geodetic effect due to the movement in 
the curved space-time around the planet) has finally been studied experimentally and the final result is in good agreement with the theory \cite{GravityPB}. The frame dragging drift was measured with four high precision electrically suspended gyroscopes mounted in the Gravity Probe B satellite in a polar circular orbit $642$ km height. Being a secular effect, it piled up orbit after orbit during one year of data collection and a final effect of $-3.72(0.72)\times 10^{-2}$ arcsec was derived. The measured effect is in good agreement with the predictions of General Relativity but the $20\%$ error is still high. It is remarkable
that this data was obtained after a careful modelling of a much larger effect on the alignment of the gyroscopes caused by
electrostatic interactions between the rotors and their housing \cite{Everitt,Keiser,Muhlfelder,Silbergleit}.

The prefactor $r_S/r$ in Eq. \ (\ref{Bgm}) is $\simeq 1.26 \times 10^{-9}$ for the Gravity Probe B orbit and this
gives us an idea of the small magnitude of the frame-dragging effect. One could think that the results of this mission
rule out the possibility of any other gravitomagnetic effect beyond conventional General Relativity. However, we must
take into account that Gravity Probe B was not designed to detect non-secular contributions. A contribution of such a kind
could arise from a gravitomagnetic field tangential, at every point, to the celestial parallel in the form:
\begin{equation}
\label{Bext}
\mathbf{B}=B(r,\theta) \boldsymbol{\hat{\phi}} \; ,
\end{equation}
where $\theta$ is the polar angle and $\phi$ is the azimuthal angle. Notice that for any circular orbit around the Earth there is
always a symmetric point, for any other point, in which the vector field in Eq.\ (\ref{Bext}) has opposite direction to the vector in the first point of the trajectory. This implies that the gyroscope precession induced by this field do not add up orbit
after orbit but cancels out. Moreover, we will calculate the contributions to the Lense-Thirring precession of the longitude of the ascending node of the geodynamics satellites to check the viability of our proposal in the context of the most accurate measurements of this effect to date\cite{IorioReview}. 

On the other hand, the effects of this new field would manifest on highly eccentric elliptical orbits or in asymmetrical flybys.
This could explain why the anomaly discovered in the flybys of the Earth only appears on these particular orbits. Our objective
is to propose a reasonable form for $B(r,\theta)$ in Eq.\ (\ref{Bext}) by dimensional and continuity arguments and analyze its potential as an explanation of the flyby anomaly. 

The paper is organized as follows: In section \ref{model} we discuss the form of the transversal gravitomagnetic field and develop a perturbative approach to analyze its influence on flybys. Section \ref{results} is devoted to the calculations of the
perturbations for the NEAR, Galileo, Cassini, Rosetta flybys. We also give a prediction for the most recent Juno flyby. The effects on the Gravity Probe B experiment and the LAGEOS, LARES and GRACE Lense-Thirring precession is discussed in Section \ref{constraints}. Finally, we remark on the context and further development of the idea of a transversal gravitomagnetic field
in Section \ref{conclusions}.
\section{Flybys in a transversal gravitomagnetic field}
\label{model}
We should begin by proposing a reasonable hypothesis for a transversal component of a gravitomagnetic field. Firstly, we can expect it to be proportional to the total angular momentum of the Earth. Moreover, the continuity condition $B(r,\theta=0)=
B(r,\theta=\pi)=0$  must be verified. We can also argue that $B(r,\theta)$ is an even function of the polar angle, $B(r,-\theta)
=B(r,\theta)$. From these conditions alone we suggest that a simple form for a transversal gravitomagnetic field is:
\begin{equation}
\label{Brt}
\mathbf{B}(r,\theta)=\beta \, \Omega  \displaystyle\frac{R}{r} \sin\theta \cos\theta \, \boldsymbol{\hat{\phi}}\; ,
\end{equation}
where $\Omega$ is the angular velocity of the Earth, $R$ is its radius and $\beta$ is a constant to be determined. The reason for the particular dependence in the polar angle, $\theta$, and the decrease as the inverse of the distance to the center of the Earth, $r$, will be revealed in Sec. \ref{constraints} in relation with the constraints imposed by other precise measurements of the Earth's gravitational field. The unit azimuthal vector, $\boldsymbol{\hat{\phi}}$ can easily be expressed in the cartesian celestial coordinate system whose $x$ axis points toward the first point of Aries (the point where the Sun crosses the celestial equator during the Vernal equinox). In this system the $z$ axis  points towards the celestial pole and this defines the remaining $y$ axis in a right handed coordinate system. So, we have:
\begin{equation}
\label{phipro}
\begin{array}{rcl}
\sin \theta \cos\phi &=& \displaystyle\frac{\mathbf{r} \cdot \ihat}{\vert \mathbf{r} \vert} \\
\noalign{\smallskip}
\sin \theta \sin\phi &=& \displaystyle\frac{\mathbf{r} \cdot \jhat}{\vert \mathbf{r} \vert} \; .
\end{array}
\end{equation}
In order to calculate the contribution of the small acceleration $\mathbf{a}=\mathbf{v}\times\mathbf{B}$ to the perturbation of the
hyperbolic orbit in a flyby we must take into account also the additional second-order effects induced by the Earth's Newtonian gravitational field, the Sun and
the Moon. We must start with some standard relations for the celestial mechanics of a flyby hyperbolic orbit. 
\subsection{Ideal flyby orbits}
\label{intro1}
\begin{figure}
\includegraphics[width=\columnwidth]{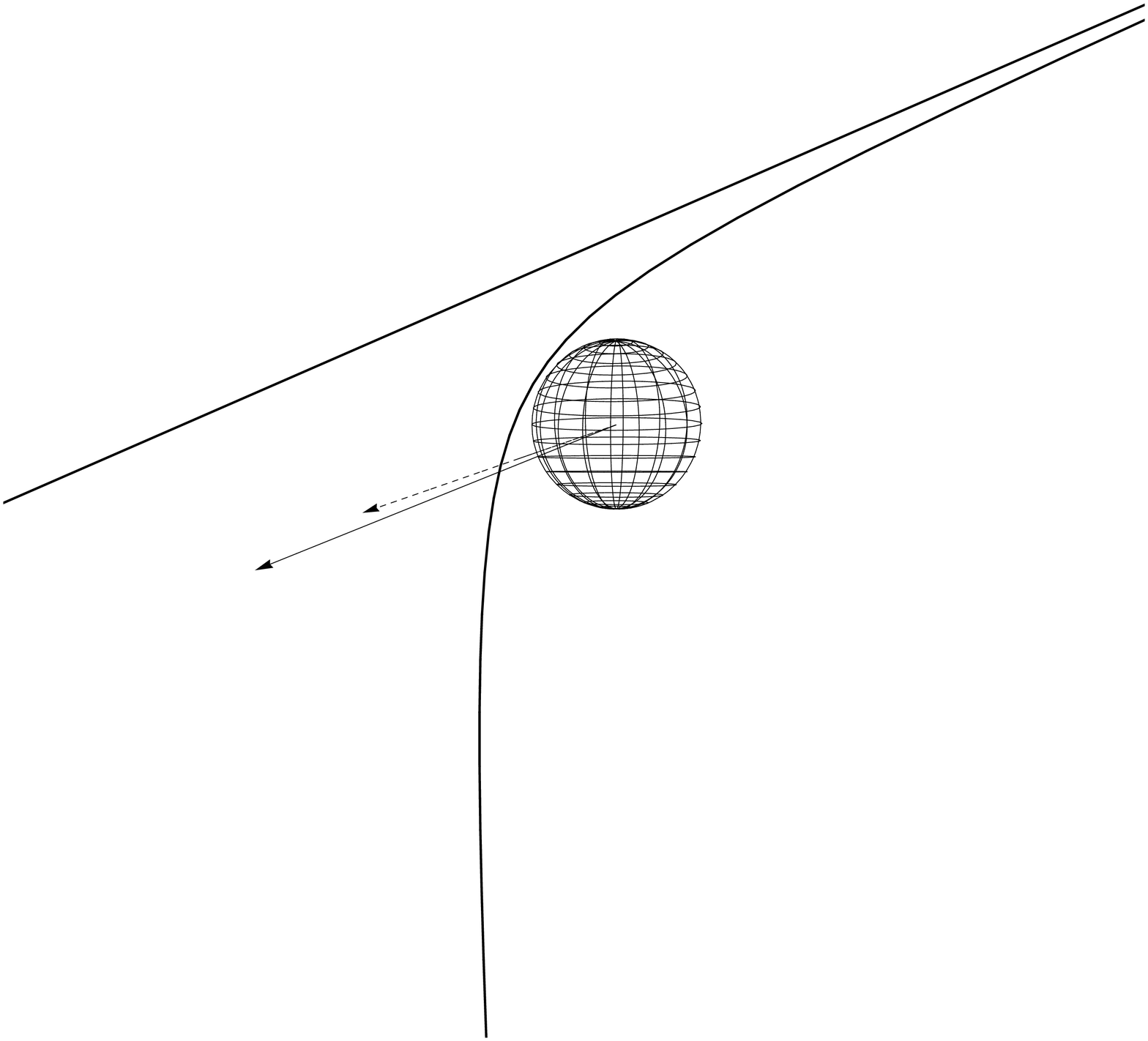}%
 \caption{\label{fig1}
Plot of the NEAR flyby orbit (January 23, 1998). The solid vector points towards the Sun and the dashed vector points towards 
the Moon at the instant of the closest approach. The asymptotic ingoing direction is also shown.}
\end{figure}

In Fig.\  \ref{fig1} we have plotted the trajectory of the NEAR spacecraft during its flyby of January 23, 1998. Parameters for this and other flyby orbits are given in Table \ref{spacedata}.
It is expedient to parametrize these hyperbolic trajectories in a coordinate system relative to the orbit. So, we choose a unit vector along the periapsis direction corresponding to the point of closest approach, $\mathbf{\hat{s}}$, a second unit vector pointing along the direction of the inclination vector of the
orbit, $\mathbf{\hat{w}}$, and a third one perpendicular to those two, $\mathbf{\hat{n}}$. This third unit vector is defined in such a way that the scalar product with the initial radiovector of the spacecraft, $\mathbf{r_{\textsc{in}}}$, is positive. These
vectors are given as follows:
\begin{equation}
\label{coordinates}
\begin{array}{rcl}
\mathbf{\hat{s}}&=& \cos \theta_p \; \khat+\sin \theta_p \cos \alpha_p \; \ihat
+\sin \theta_p \sin \alpha_p \; \jhat\\
\noalign{\smallskip}
\mathbf{\hat{w}}&=& \cos I\; \khat+\sin I \cos \alpha_I\; \ihat
+\sin I \sin \alpha_I\; \jhat \\
\noalign{\smallskip}
\mathbf{\hat{n}}&=&\pm \mathbf{\hat{w}} \times \mathbf{\hat{s}}\; ,
\end{array}
\end{equation}
where $\theta_p$, $\alpha_p$ are the celestial polar angle and right ascension of the periapsis, $I$ and $\alpha_I$ are the inclination and the right ascension of the inclination vector and the sign in the last expression for $\mathbf{\hat{n}}$ depends
on the orientation of the orbit. The orthogonal system $\ihat$, $\jhat$, $\khat$ is, obviously, the celestial coordinate system.

In the NEAR flyby case the parameters were (all angles in degrees):
$\theta_p=57$, $\alpha_p=280.43$, $I=108.0$, $\alpha_I=\alpha_p+\arccos(-\cot I\cot \theta_p)=358.24$. The incoming direction is given by $\theta_i=69.24$ and $\alpha_i=81.17$. In this particular case, it can be shown that $\mathbf{\hat{n}}=\mathbf{\hat{w}}\times \mathbf{\hat{s}}$.

It is also usual to use the eccentric anomaly, $\eta$, instead of time, $t$, or the true anomaly, $\nu$. The true anomaly is the angle formed by the radiovector at any instant and the radiovector at the instant of closest approach to the Earth. The relation among $\eta$ and $\nu$ is given by:
\begin{equation}
\label{eccanomaly}
\cosh \eta=\displaystyle\frac{\epsilon+\cos \nu}{1+\epsilon \cos \nu}\; ,
\end{equation}
where $\epsilon > 1$ the eccentricity of the hyperbolic orbit. The time of flight can also be given in terms of the eccentric anomaly by:
\begin{equation}
\label{timeofflight}
t=T (\epsilon \sinh \eta -\eta)\; ,
\end{equation}
where the time-scale $T=\sqrt{ (-a)^3/\mu}$, $a$ being the semi-major axis of the orbit and $\mu$ is the product of the  gravitational constant and the mass of the Earth, $\mu_{\textsc{E}}=398600.4\, \mbox{km}^3/\mbox{s}^2$. 

The equations for the radiovector and the velocity of the spacecraft in the ideal hyperbolic orbit are then given by
\begin{eqnarray}
\label{rhyper}
\mathbf{r}(\eta)&=&a (\cosh \eta-\epsilon)\, \mathbf{\hat{s}}-a \sqrt{\epsilon^2-1} \sinh \eta\, \mathbf{\hat{n}}\\
\noalign{\smallskip}
\label{vhyper}
\mathbf{v}(\eta)&=&\displaystyle\frac{a}{T(\epsilon \cosh \eta-1)}(\sinh \eta \, \mathbf{\hat{s}} \nonumber \\
\noalign{\smallskip}
&-&\sqrt{\epsilon^2-1} \cosh \eta\, \mathbf{\hat{n}})\; ,
\end{eqnarray}
Concerning the tracking of the spacecraft and the inference of the flyby orbit from the tracking data there are many
practical problems still not analyzed in full detail. An accurate flyby model should take into account the most recent
GGM03 model for the Earth gravitational field as elaborated from the data of the GRACE satellite \cite{GRACEmod}. This model
has achieved a large geographical resolution and a precision on the variations of the gravitational field at the Earth's surface down to 1 milligal ($10^{-5}$ m s$^{-2}$). On the other hand, the position and velocity of the spacecraft is tracked by 
different ground stations around the globe in different relative movements with respect to the spacecraft. This should be also
be taken into account in the modelling of the Doppler effect. Preliminary assessments showed that these effects are relatively
small and unimportant in connection with the anomaly \cite{Lammerzahl,STEQUEST} but their rigorous analysis should help in the improvement of the error bars on the anomalous accelerations. 
\subsection{First and second-order perturbations}
\label{intro2}
In the following it is useful to use the non-dimensional parameters, $\tau=t/T$ for time and $ \mathbf{v}^\star=T\delta
\mathbf{v}/\vert a \vert$ for the velocity. A first-order perturbation comes from the acceleration imparted by the
gravitomagnetic field in Eq.\ (\ref{Brt}) as follows:
\begin{equation}
\label{amg}
\mathbf{a}_{\textsc{mg}}=B^\star(r,\theta)\,\mathbf{v}^\star \times \boldsymbol{\hat{\phi}}\; ,
\end{equation}
with $B^\star(r,\theta)=B(r,\theta)*T$. The cross product in Eq. \ (\ref{amg}) is readily calculated in cartesian
coordinates from Eqs.\ (\ref{phipro}), (\ref{coordinates}) and (\ref{vhyper}). The position of the spacecraft deviates
from the ideal hyperbolic orbit discussed in Section \ref{intro1} as a consequence of this first-order acceleration. However, this implies another second-order contribution from the change of the Earth's Newtonian gravitational field at the new
orbital points:
\begin{equation}
\label{aEarth}
\delta\mathbf{a}_{\textsc{Earth}}=-\mu_{\textsc{E}}^\star\, \displaystyle\frac{\delta \mathbf{r}^\star}{r^{\star 3}}+3\,
\mu_{\textsc{E}}^\star\, \mathbf{r}^\star \cdot \delta\mathbf{r}^\star \displaystyle\frac{\mathbf{r}^\star}{r^{\star 5}}\; , 
\end{equation}
where $\mu_{\textsc{E}}^\star=T \mu_{\textsc{E}}/\vert a \vert^3$ is the scaled gravitational constant times the mass
of the Earth. As it is well-known, most of the orbital perturbations come from the tidal forces exerted by the Sun (and in 
minor proportion by the Moon). If the Sun is located at radiovector $\mathbf{R}$, with the center of the Earth as origin, and the spacecraft is found at $\mathbf{r}$ the tidal force is given by:
\begin{equation}
\label{atidal}
\mathbf{a}_{\textsc{tidal}}=\mu_{\textsc{S}}\left(-\displaystyle\frac{\mathbf{R}}{R^3}+\displaystyle\frac{\mathbf{R}-\mathbf{r}}{\left(r^2+R^2-2 \mathbf{r}\cdot\mathbf{R}\right)^{3/2}} \right)\; ,
\end{equation}
with $\mu_{\textsc{S}}=1.3271244\times10^{11} \mbox{km}^3/\mbox{s}^2$ as the product of the gravitational constant and the mass
of the Sun. The change in the position of the spacecraft from the transversal gravitomagnetic force induces a second-order contribution from the tidal force as follows:
\begin{equation}
\begin{array}{rcl}
\label{aSun}
\delta \mathbf{a}_{\textsc{Sun}}&=&-\mu_{\textsc{S}} \displaystyle\frac{\delta \mathbf{r}}{\left(r^2+R^2-2\mathbf{r}\cdot\mathbf{R}\right)^{3/2}}\\
\noalign{\smallskip}
&+&3\mu_{\textsc{S}} \displaystyle\frac{(\mathbf{R}-\mathbf{r})\cdot
\delta \mathbf{r}}{\left(r^2+R^2-2\mathbf{r}\cdot\mathbf{R}\right)^{5/2}} \left(\mathbf{R}-\mathbf{r}\right) \; .
\end{array}
\end{equation} 
Assembling these results in Eqs.\ (\ref{amg}),(\ref{aEarth}) and (\ref{aSun}) we find that the evolution of the perturbations
in the position and the velocity of the spacecraft coming from the effect of the transversal gravitomagnetic field should
be calculated from the system:
\begin{equation}
\label{system}
\begin{array}{rcl}
\displaystyle\frac{d \delta \mathbf{r}^\star}{d\tau}&=&\delta \mathbf{v}^\star \\
\noalign{\smallskip}
\displaystyle\frac{d \delta \mathbf{v}^\star}{d\tau}&=&\mathbf{a}_{\textsc{mg}}+\mathbf{a}_{\textsc{Earth}}+
\mathbf{a}_{\textsc{Sun}} \; .
\end{array}
\end{equation} 
By solving this system for the differential equations of motion we should be able to compare with the anomalous perturbations
of the Earth flybys in the next section. To perform this integration it is usually more convenient to employ the eccentric anomaly instead of time as defined in Eq. (\ref{timeofflight}).
\section{Numerical results}
\label{results}
In Table \ref{spacedata} we show the parameters for the orbits of several spacecraft flybys in which the anomalous variation of the velocity has been found. We have displayed only the parameters necessary for the calculations including the orbital eccentricity, $\epsilon$, the semi-major axis of the hyperbolic orbit, $a$, the polar and azimuthal angles locating the incoming and outgoing directions, the perigee and the inclination vector. These data were collected or inferred from the JPL Horizons on-line system data \cite{Horizons} as well as Anderson et al. paper \cite{FlybyPRL}. We should also know the average
position of the Sun during the flyby as this is the major source of second-order perturbations. These data are tabulated in 
Table \ref{table2} for the eight flybys considered.
\begin{table*}
\caption{Parameters for the spacecraft flybys of the Earth.}
\centering
 \begin{tabular}{lcccccccccc}\label{spacedata}
  Spacecraft & Date & $\epsilon$ & $a$ (km) & $\theta_{\textsc{in}}$ & $\theta_{\textsc{out}}$ & $\theta_{\textsc{p}}$ 
  & $I$ & $\alpha_{\textsc{in}}$ & $\alpha_{\textsc{p}}$ & $\alpha_{\textsc{I}}$ \\
    \noalign{\smallskip}
  NEAR & 1/23/1998 & 1.8135 & -8494.87 & 69.24$^\circ$ & 161.96$^\circ$ & 57$^\circ$ & 108$^\circ$ & 81.17$^\circ$ &
  280.43$^\circ$ & 358.25$^\circ$ \\
  \noalign{\smallskip}
  Galileo I & 12/8/1990 & 2.4729 & -4977.24 & 77.48$^\circ$ & 124.25$^\circ$ & 64.8$^\circ$ & 142.9$^\circ$ & 86.60$^\circ$
  & 319.96$^\circ$ & 11.48$^\circ$ \\
  \noalign{\smallskip}
  Galileo II & 12/8/1992 & 2.3194 & -5058.31 & 55.74$^\circ$ & 94.87$^\circ$ & 123.8$^\circ$ & 138.7$^\circ$ & 39.47$^\circ$
  & 302.72$^\circ$ & 77.56$^\circ$ \\
  \noalign{\smallskip} 
  Cassini & 8/18/1999 & 5.8525 & -1555.09 & 102.92$^\circ$ & 94.99$^\circ$ & 113.5$^\circ$ & 25.4$^\circ$ & 154.33$^\circ$ & 245.59$^\circ$ & 221.90$^\circ$ \\
  \noalign{\smallskip}
  Rosetta & 3/4/2005 & 1.3118 & -26710.9 & 92.81$^\circ$ & 124.29$^\circ$ & 69.8$^\circ$ & 144.9$^\circ$ & 166.68$^\circ$ & 22.71$^\circ$ & 324.28$^\circ$ \\
  Rosetta II & 11/13/2007 & 1.5401 & -33417.5 & 79.32$^\circ$ & 71.70$^\circ$ & 154.7$^\circ$ & 115.0$^\circ$ & 45.95$^\circ$
  & 304.0$^\circ$ & 130.9$^\circ$ \\
  \noalign{\smallskip}
  Rosetta III & 11/13/2009 & 1.5976 & -25491.1 & 108.4$^\circ$ & 65.65$^\circ$ & 97.44$^\circ$ & 155.6$^\circ$ & 31.78$^\circ$
  & 276.2$^\circ$ & 169.5$^\circ$ \\
  \noalign{\smallskip}
   Juno & 9/10/2013 & 4.6489 & -3645.92 & 104.21$^\circ$ & 50.59$^\circ$ & 123.39$^\circ$ & 47.13$^\circ$ & 215.40$^\circ$ & 344.14$^\circ$ & 291.85$^\circ$  
  \end{tabular}
\end{table*}

\begin{figure}
\includegraphics[width=\columnwidth]{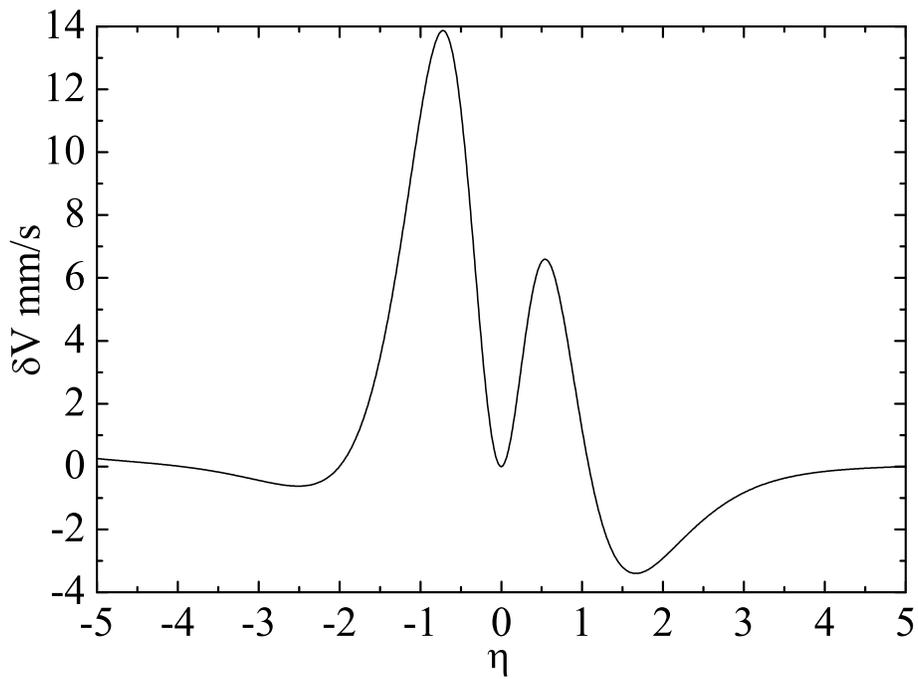}%
\caption{\label{fig2}
Velocity perturbations induced by the transversal gravitomagnetic field in the Galileo II flyby. In this case we have
taken $\beta=2.0\times 10^{-3}$. Notice that the average difference between the sections of the orbits after and before
the perigee is negative in agreement with the observed decrease during this flyby.}
\end{figure}

\begin{figure}
\includegraphics[width=\columnwidth]{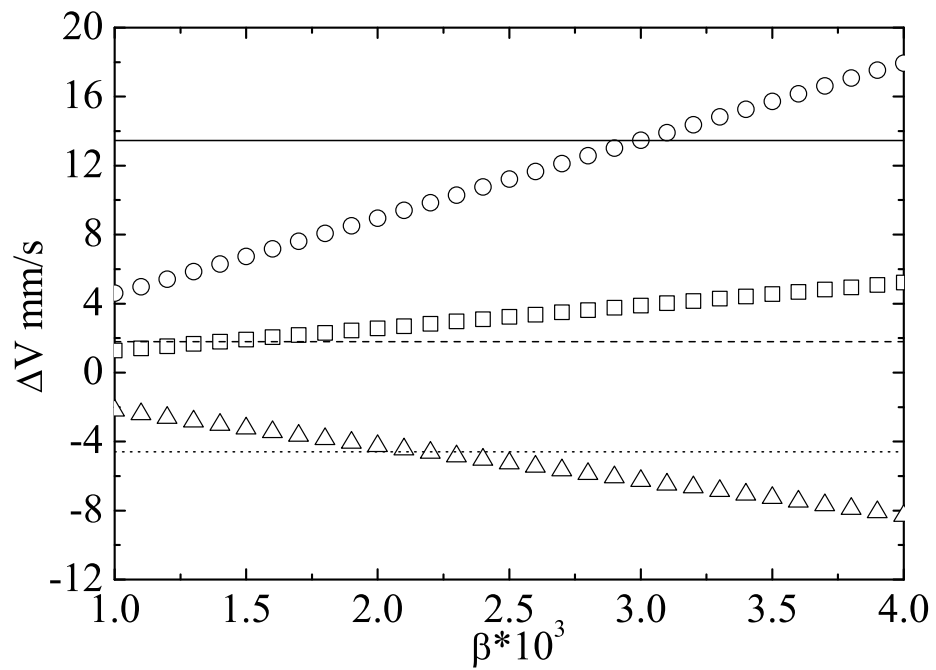}%
\caption{\label{fig3}
Modelled anomalous increases of the velocity during the NEAR (circles), Rosetta (squares) and the Galileo II flybys (triangles) versus the parameter $\beta$. Horizontal solid, dashed and dotted lines correspond to the observed values for the NEAR, Rosetta and Galileo II flybys, respectively.}
\end{figure}

\begin{figure}
\includegraphics[width=\columnwidth]{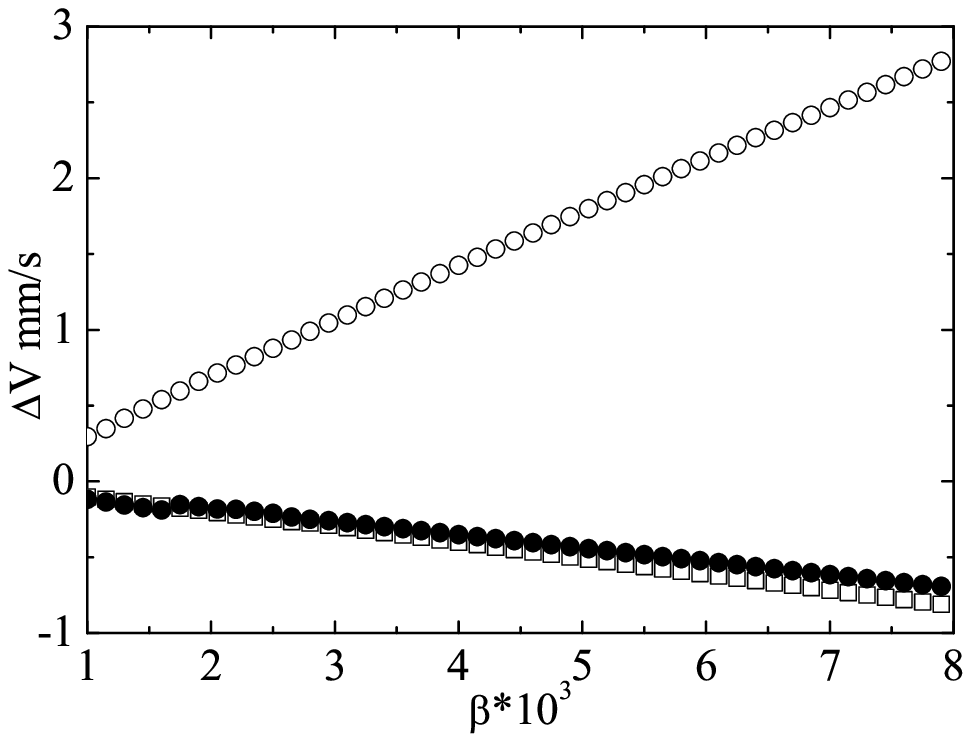}%
\caption{\label{fig4}
The same as Fig. \protect\ref{fig3} but for the Galileo I (open circles), Juno (closed circles) and Cassini (squares) flybys. In the case of the Galileo I flyby a velocity increase of $3.92$ mm/s was reported. This was a decrease of $-2$ mm/s for the Cassini spacecraft. Notice that, according to this model, the result would be similar for the most recent Juno flyby.}
\end{figure}

\begin{table}
\centering
\caption{Average distance and average unit vector towards the Sun during the flybys in the celestial cartesian coordinate system.}
\begin{tabular}{lcc}
\label{table2}
Spacecraft & $\langle R_{\textsc{S}} \rangle$ ($10^8$ km) & $\langle \mathbf{\hat{r}}_{\textsc{S}} \rangle$ \\
\noalign{\smallskip}
NEAR & 1.4727 & (0.5413,-0.7700,-0.3338)\\
\noalign{\smallskip}
Galileo I & 1.4739 & (-0.2594,-0.8852,-0.3838) \\
\noalign{\smallskip}
Galileo II & 1.4737 & (-0.2510,-0.8872,-0.3847) \\
\noalign{\smallskip}
Cassini & 1.5147 & (-0.8072,0.5403,0.2343) \\
\noalign{\smallskip}
Rosetta & 1.4835 & (0.9587,-0.2580,-0.1119) \\
\noalign{\smallskip}
Rosetta II & 1.4809 & (-0.6513,-0.6951,-0.3013) \\
\noalign{\smallskip}
Rosetta III & 1.4808 & (-0.6447,-0.7002,-0.3035) \\
\noalign{\smallskip}
Juno & 1.4882 & (-0.9615,-0.2479,-0.1075) 
\end{tabular}
\end{table}
Using Eqs. (\ref{amg}), (\ref{aEarth}) and (\ref{aSun})we have performed a numerical integration of the system in Eq. (\ref{system}) for $\eta=0$ onwards and backwards. A typical result for the Galileo II flyby is plotted in Fig. \ref{fig2}.
In this particular case we see that the perturbation corresponds to a decrease of the orbital velocity after the closest approach to the Earth. The observed variation in the velocity can be compared with the differences in the temporal averages of $\delta V$
after and before this perigee. A simple rough measure of the variation $\delta V$ can be obtained as the difference among the peak perturbations occurring in the post-encounter and pre-encounter regions. In Fig.\ (\ref{fig3}) we have plotted the results for the variation of $\delta V$ in the cases of the NEAR, Rosetta and Galileo II flybys in terms of the value of the parameter $\beta$ in Eq.\ (\ref{Bext}). In the NEAR case the largest velocity increase, amounting to $13.46$ mm/s as it was deduced from the X-band Doppler residuals, was detected \cite{FlybyPRL}. This amounted only to $1.8$ mm/s for the Rosetta flyby.
The Galileo II spacecraft approached sufficiently close to the Earth surface (a $303$ km altitude at the perigee) for the atmospheric drag effect to be important. Nevertheless, Anderson
et al. deduced an unexpected decrease of $-4.6$ mm/s. These results are compatible with a $\beta$ parameter value in the range $1.4\times 10^{-3}\le \beta \le 3.0 \times 10^{-3}$.
Similarly, in Fig. (\ref{fig4}) we plot the results for the Galileo I, Juno and Cassini flybys. For this Cassini flyby (whose perigee was much higher for any atmospheric drag to be relevant) a decrease of $-2$ mm/s was derived from the orbital data \cite{FlybyPRL}. 
Our model can also predict the result for the still not analyzed Juno flyby. The prediction is a behaviour very similar to the Cassini flyby and, consequently, a similar decrease is expected. On the contrary, Anderson et al.'s fitting equation \cite{FlybyPRL,IorioJuno} predicts an increase around $6$ mm/s. However, we must consider that the uncertainties could be large in these isolated measurements. 

Finally, we must discuss the Rosetta II and Rosetta III flybys, which have been the subject of some polemic \cite{STEQUEST}. 
The phenomenological formula by Anderson et al. in Eq. (\ref{flybyform}) predicts small, but measurable, increases of the asymptotic velocity around $0.36$ mm/s and $0.46$ mm/s for the Rosetta II and III, respectively. However, a careful analysis of the data showed no anomalous behaviour of the fitted trajectory. Consequently, despite the success in fitting the previous flyby
anomalies, we cannot attribute a general validity to that equation. Using the gravitomagnetic model discussed in this paper we obtain $\Delta V \simeq 0.65$ mm/s for $\beta=10^{-3}$ in the last Rosetta flyby, in the same range of te prediction of Eq. (\ref{flybyform}), but different from the zero result obtained from the data analysis.

We have also noticed that the value of $\beta$ that we need to fit the anomaly in the case of the Cassini and Galileo I flybys is
larger than the one corresponding to the other three flybys (NEAR, Galileo II and Rosetta). Inspecting Table \ref{spacedata} we can associate this inconsistency in the $\beta$ values with the larger eccentricities in the Cassini and Galileo I orbits. On the other hand, we could elaborate an improved model in which $\beta$ is not a constant but a function of $r$ in order to obtain a
better fit of the flybys. However, we are about to see that further speculation along this line is not required because the transversal gravitomagnetism phenomenological model cannot be made compatible with the flyby anomalies as well as the Gravity Probe B and the geodynamics satellites data.
\section{Observational Constraints}
\label{constraints}
A problem which arises in any putative model of the flyby anomaly is that we have already many constraints based upon careful measurements of the Earth's gravitational field by
using low orbit satellites. The extra perturbation detected in the flybys can be considered very large in the context of the present degree of accuracy achieved in these measurements.
Consequently, we must discuss the contributions of any new interaction, postulated to explain the flyby anomaly, in other experiments performed on satellites and check whether its contribution
is sufficiently small to remain still undetected or not. There are two high-precision measurements of the Earth's gravitational field (including gravitomagnetic effects): the geodynamics satellites and
the Gravity Probe B experiment. 

The LAGEOS and LAGEOS II are spherical artificial spacecraft entirely covered by retroreflectors which have been accurately tracked by laser ranging since their launch in 1976 and 1992, respectively.
One of the objectives of these geodynamics satellites is to check the so-called Lense-Thirring effect, i. e., the precession of the longitude of the ascending node of the orbit as the consequence of the perturbation induced by the Earth's gravitomagnetic field as predicted by General Relativity. This effect is very small (around 30 milliarcsec/year) in comparison with the much larger perturbations contributed by the zonal harmonic coefficients of the Earth's geoid \cite{Ciufolini}. However, the classical nodal precession depends on the cosine of the inclination of the orbit in contrast with the relativistic Lense-Thirring effect which is independent of the orbital plane orientation. This gives a chance to deduce the tiny relativistic effect by comparing the data for several satellites. More recently, the similar LARES and GRACE missions added more data to the analysis. Despite these efforts, no conclusive results have been obtained yet \cite{IorioReview}. 

We should now calculate the contribution of the postulated extra gravitomagnetic field in Eq.\ (\ref{Brt}) to the precession of the longitude of the ascending node in these artificial satellites. In terms of the true anomaly, $\nu$, the equations of the elliptic orbit can be written as follows \cite{Valtonen}:
\begin{eqnarray}
\label{orbit}
\mathbf{r}&=&p \displaystyle\frac{ \cos \nu\, \hat{\mathbf{n}}+\sin\nu\, \hat{\mathbf{s}}}{1+\epsilon \cos(\nu-\omega)} \\
\noalign{\smallskip}
\mathbf{v}&=& \displaystyle\sqrt{\displaystyle\frac{\mu_{\textsc{E}}}{p}} \left\{ -(\sin\nu+\epsilon \sin\omega)\, \hat{\mathbf{n}} \right.\\
\noalign{\smallskip}
&+& \left. (\cos\nu+\epsilon\cos\omega)\, \hat{\mathbf{s}} \right\} \; ,
\end{eqnarray}
where $p=a (1-\epsilon^2)$ is the semi-latus rectum and $\omega$ is the argument of the pericenter, i. e., the angle between the radiovector of the periapsis and the direction of the ascending node. The system of
reference in cartesian coordinates is given by $\hat{\mathbf{n}}=\ihat$, $\hat{\mathbf{s}}=\cos I \, \jhat+ \sin I \, \khat$ and 
$\hat{\mathbf{w}}=\cos I\, \khat-\sin I\, \jhat$, $I$ being the inclination of the orbital plane. So $\hat{\mathbf{n}}$ points towards the direction of the ascending node, $\hat{\mathbf{w}}$ is the 
inclination vector and $\hat{\mathbf{s}}$ forms a right-handed orthogonal system with them. 

The polar and azimuthal angles can also be expressed in terms of the true anomaly and the inclination of the orbit as follows:
\begin{eqnarray}
\label{angles}
\cos \theta&=&\sin \nu \sin I \\
\noalign{\smallskip}
\cos \phi&=&\displaystyle\frac{\cos \nu}{\sqrt{1-\sin^2 I \sin^2 \nu}}\; .
\end{eqnarray}
The contribution of the perturbation forces to the instantaneous variation of the longitude of the ascending node, $\Omega$, is given by \cite{Burns,Pollard,Danby}:
\begin{equation}
\label{Omegapert}
\displaystyle\frac{d \Omega}{d t} = \displaystyle\frac{r \sin\nu}{H \sin I}\, \mathcal{N} \; ,
\end{equation}
where $H=\sqrt{\mu_{\textsc{E}} p}$ is the angular momentum of the orbiting spacecraft per unit mass and $\mathcal{N}$ is the
component of the perturbing force perpendicular to the orbital plane. In our case, we have:
\begin{equation}
\label{FNormal}
\mathcal{N}=\beta \, \Omega \, \displaystyle\frac{R}{r} \, \sin\theta \cos\theta\, \mathbf{v} \times \hat{\mathbf{\phi}} \cdot \hat{\mathbf{w}} \; ,
\end{equation}
From Eqs. (\ref{orbit}),({\ref{angles}) and (\ref{Omegapert}) we finally have:
\begin{equation}
\label{dOmegadt}
\begin{array}{rcl}
\displaystyle\frac{d \Omega}{d t}&=&\beta \, \Omega \, \displaystyle\frac{R}{a (1-\epsilon^2)}  \left\{ \sin^3 \nu \cos\nu (\sin I-\cos I) \right.\\
\noalign{\smallskip}
 &+&\left. \epsilon \cos \omega \sin^3 \nu \sin I-\epsilon \sin \omega \sin^2 \nu \cos \nu \cos I \right\}\; .
\end{array}
\end{equation}
Remarkably, every term in Eq. (\ref{dOmegadt}) cancels out when averaged over a full orbit, $\nu \in (0, 2 \pi)$. As the determination of the Lense-Thirring precession is plagued with difficulties, arising from the very large contributions of the 
Earth's zonal harmonic coefficients \cite{IorioReview,Ciufolini}, it could be argued that a phenomenological law, as that in Eq. (\ref{Brt}), could consistently fit the flyby anomalies without impacting noticeably in the precession of the line of nodes of the satellites. However, if a sufficiently strong field of the form in Eq.\ (\ref{Brt}) is assumed, the contribution to the precession of a gyroscope in a low orbit exhibits large oscillations during the orbit which should have been found in the Gravity Probe B experiment \cite{GravityPB}. The result shown in Fig.\ \ref{fig5} implies a cancellation of the precession every half-orbit. Nevertheless, the average value of these oscillations is as large as $11.6$ arcsec for the $\beta=2 \times 10^{-3}$ parameter which we have shown to be compatible with the flyby anomalies. 

It is also interesting to provide an estimation of the
value of the bounds on the value of $\beta$ according to the Gravity Probe B experiment. The North-South orientation of the gyroscopes was measured with the average precision of $18.3\times 10^{-3}$ arcsec/yr in connection with the determination of the geodetic effect which implied a secular reorientation of the gyroscopes in that direction \cite{GravityPB}. Consequently, we can estimate a maximum non-secular average fluctuation of the North-South orientation of the gyroscopes of this magnitude, which should correspond to $\beta \le 3 \times 10^{-6}$. Clearly insufficient to explain the flyby anomaly but, perhaps, within the range of other measurements in the Solar System. 
\begin{figure}
\includegraphics[width=\columnwidth]{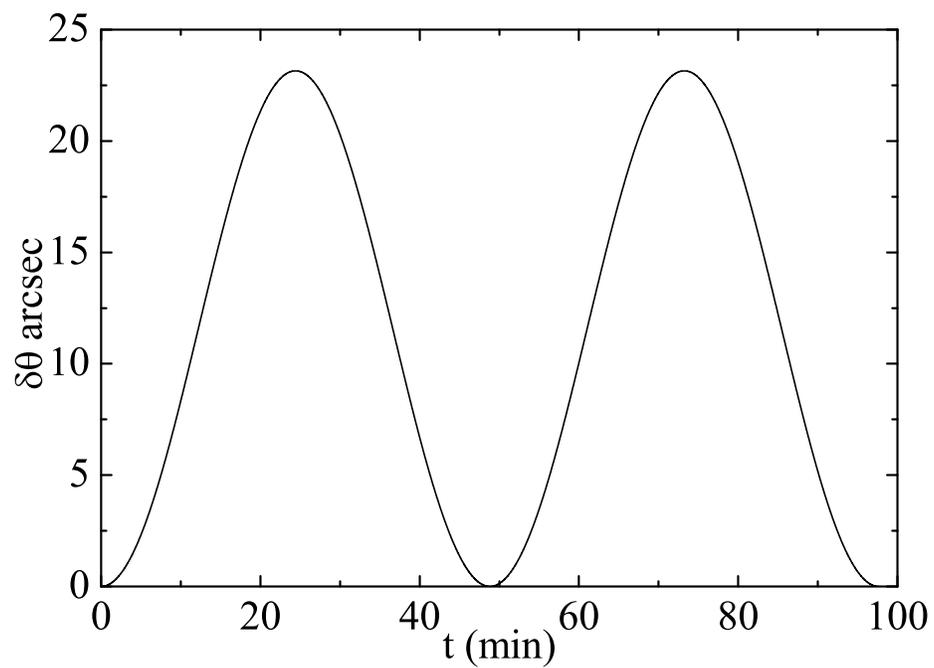}%
\caption{\label{fig5}
Contribution to the North-South orientation of a gyroscope in the Gravity Probe B orbit by a gravitomagnetic field of the form given in Eq.\ (\protect\ref{Brt}) with $\beta=2 \times 10^{-3}$.} 
\end{figure}

\section{Conclusions and Remarks}
\label{conclusions}
In this paper we have explored the idea of a strong gravitomagnetic field as the source of the flyby anomaly. Although this has
been done in a pure phenomenological context is not at all excluded in principle.  The condition of  general covariance is too weak to restrict the form of the field equations of General Relativity, as many authors have recognized since the early days of the theory almost a century ago. Since them, many alternatives have been considered and dismissed \cite{Goenner} and some with local and cosmological implications, such as torsion models, are still a field of active research \cite{Hehl,Poplawski}. We have shown that a transversal component of a gravitomagnetic field would exhibit a maximum contribution to orbital perturbation effects in hyperbolic or highly eccentric elliptical orbits and it can be defined in such a way that the secular contribution to the Lense-Thirring precession is null. This is, precisely, what in the general features of the flyby anomaly is found.
 
On the other hand, the detailed orbital evolution of the perturbations in gyroscope and Lense-Thirring precessions appear to be 
very large for a sufficiently strong field predicting a velocity variation in the millimeter range among the post-encounter and pre-encounter sections of the hyperbolic flyby orbit. Similar stringent constraints have been found in the earth-bound dark matter model \cite{haloDM}. The difficulties in settling interaction models verifying the conditions of being compatible with precise measurements of satellites in circular and elliptical orbits and also the large anomalies in the flyby data could point
out towards a possible systematic error as the source of the discrepancies in the X-band Doppler residuals \cite{FlybyPRL,Acedo}.

Anyway, many theoretical and computational work remains to be done in order to clarify all the conventional sources of perturbation on the flyby trajectories. In particular, a detailed numerical model of the atmospheric drag taking into account the density of the atmosphere and the geometry of the spacecraft in a finite-element model is necessary to 
provide any reliable prediction of the dragging effects suffered by the spacecraft crossing the thermosphere during their closest
approach to the Earth (this was specially noticeable in the case of the Galileo II flyby). The influence of albedo and infrared radiation will also be possible to analyze in detailed models of the spacecraft geometry as those build for the Pioneer 10 and 11
\cite{PioneerI,PioneerII}.

The solution to this riddle could also be found through the forthcoming Space-Time Explorer and Quantum Equivalence Principle Space Test (STE-QUEST) mission in which a spacecraft would be launched to a an elliptic orbit with very large eccentricity $\epsilon=0.773$ specially suited to simulate an Earth flyby every period of $16$ hours \cite{STEQUEST}. Another forthcoming mission: Gravity Recovery and Climate Experiment-Follow On (GRACE-FO) will also provide even more accurate data on the Earth's gravitational field that could be instrumental in the next few year to elucidate this issue \cite{GRACEFO}.

\bibliographystyle{model2-names}


%
%

\end{document}